\documentstyle[preprint,prl,aps]{revtex}
\begin{document}
\def\sn2{$\sin^22\theta$}
\def\dm2{$\Delta m^2$}
\def\ch2{$\chi^2$}
\draft
\begin{titlepage}
\preprint{\vbox{\baselineskip 15pt{\hbox{IFP-480-UNC}
\hbox{Ref. SISSA 177/93/EP}
\hbox{hep-ph 9311306} \hbox{November 1993}
\hbox{(Updated Version, March 1994)}}}}
\title{ \large \bf New Constraints on Neutrino Oscillations in Vacuum
as a Possible Solution of the Solar Neutrino Problem }
\author{P. I. Krastev\footnote{Permanent address:
Institute of Nuclear Research and Nuclear Energy, Bulgarian Academy
of Sciences, BG--1784 Sofia, Bulgaria.}}
\address{Institute of
Field Physics, Department of Physics and Astronomy, \\
The University
of North Carolina at Chapel Hill, CB -- 3255, Phillips Hall, \\
Chapel Hill, NC 27599-3255}
\author{S.T. Petcov$^*$}
\address{Scuola Internazionale Superiore di Studi Avanzati, and \\
Instituto Nazionale di Fizica Nucleare, Sezione di Trieste, \\
Via Beirut 2 -- 4, I-34013 Trieste, Italy}
\maketitle
\begin{abstract}
\begin{minipage}{5in}
\baselineskip 16pt
 Two-neutrino oscillations in vacuum are studied as a possible
solution of the solar neutrino problem. New constraints on the
parameter \sn2, characterizing the mixing of the electron neutrino
with another active or sterile neutrino, as well as on the
mass--squared difference, \dm2, of their massive neutrino components,
are derived using the latest results from the four solar neutrino
experiments. Oscillations into a sterile neutrino are ruled out at
99\% C.L. by the observed mean event rates even if one includes the
uncertainties of the standard solar model predictions in the analysis.
\end{minipage}
\end{abstract}
\end{titlepage}
\newpage

  Neutrino oscillations in vacuum\cite{pontecorvo1},\cite{maki} have
been discussed in connection with the solar neutrino experiments
\cite{pontecorvo2} and as a possible
solution of the solar neutrino problem\cite{bahc} for about 26
years\cite{bilpont}. In this scenario it is assumed that the state
vector of the electron neutrino produced in the center of the Sun is a
coherent superposition of the state vectors of neutrinos having
definite but different masses. The flavor content of the neutrino
state vector changes periodically between the Sun and the Earth due to
the different time evolution of the vector's massive neutrino
components. The neutrinos that are being detected in the solar
neutrino detectors on Earth are thus in states representing, in
general, certain superpositions of the states of $\nu_e$, $\nu_{\mu}$,
$\nu_{\tau}$, and $\nu_s$, the latter being sterile neutrino. As the
muon, tau and sterile neutrinos interact weaker with matter than
electron neutrinos, the measured signals should be depleted with
respect to the expected ones.  This could explain the solar neutrino
problem.

  In order to make specific predictions for the signals in the solar
neutrino detectors one should i) average over the neutrino production
region in the Sun, ii) take into account the changing distance between
the Earth and the Sun\cite{grib}, and iii) integrate over the neutrino
energy spectrum\cite{bacfrau}. Qualitatively, the required depletion
of the solar $\nu_e$ flux can take place only if
\cite{bilpont} the neutrino oscillation
length in vacuum, $L_v$, is of the order of the distance between the
Earth and the Sun, $L_{se}$: for $L_v \gg L_{se}$ there will be no
time for the oscillations to develop, and in the opposite case, $L_v
\ll L_{se}$, the depletion is neutrino energy independent and can be
at most $1/N_f$, where $N_f$ is the number of weak eigenstate
neutrinos taking part in the oscillations. Despite this "tuning"
problem which has been addressed in several papers\cite{glkraus} and
which is absent in the case of the MSW solution\cite{msw} of the solar
neutrino problem, the vacuum oscillations provide an attractive
explanation of the solar neutrino observations which should be further
tested experimentally.

   Analyses of solar neutrino data in terms of two--neutrino
oscillations in vacuum have been made previously by several
groups\cite{barg}, \cite{acker}, \cite{kp}, \cite{kern}.  It was found
that a small region of values of the two parameters \dm2 and \sn2
characterizing the oscillations, \dm2~$\cong (0.55~\div~1.1) \times
10^{-10}~{\rm eV}^2$ and \sn2 $\cong (0.75 \div 1.0)$
\cite{kp}, is allowed by the data.
After the studies [10] -- [13] were completed new data have been
accumulated and published by three of the four operating experiments.

   In this letter the results of a joint analysis of the available
data (including the latest results) from all solar neutrino
experiments are presented. The first one is the pioneer Cl--Ar
experiment by R. Davis and his group\cite{clar}. Data from 84 runs of
measurements performed between 1971 and 1991 are available from this
experiment. The first and the last day of the data taking period for
each individual run have been taken from the table published
in\cite{lande}. The second experiment is the one conducted by the
Kamiokande collaborations which published recently their latest
average results \cite{kam}. Data from 13 separate ``runs'', taken
between 1986 and 1990 are also available. Each ``run'' is three months
long, only the last being slightly longer. The third experiment is the
Ga--Ge experiment conducted by the SAGE collaboration at the Baksan
Neutrino Laboratory. Only the last result\cite{sage} for the mean
value of the $^{71} {\rm Ge}$ production rate has been used by
us. Finally, the fourth experiment is the Ga--Ge one conducted by the
GALLEX collaboration, from which new data from 30 accomplished runs
have become available recently\cite{gallex}.

  The data analysis has been performed in two different ways. First,
we perform a $\chi^2-$analysis of the mean values of the event rates
in each of the four detectors. We compare the event rates expected,
assuming neutrino oscillations in vacuum take place between the Sun
and the Earth, with the experimentally measured event rates. The
expected event rates without oscillations, as well as the spectra of
the different components (pp, $^{7}{\rm Be}$, $^{8}{\rm B}$, etc.) of
the solar neutrino flux have been taken from\cite{bp}. For the ratios
of measured to expected event rates in each solar neutrino detector
the following mean values and their corresponding error bars have been
used:
\begin{equation}
R_{Cl} = 0.29 \pm 0.03,
\end{equation}
%\newpage
\begin{equation}
R_{\rm SAGE} = 0.53 \pm 0.19,
\end{equation}
\begin{equation}
R_{\rm GALLEX} = 0.60 \pm 0.09,
\end{equation}
\begin{equation}
R_{\rm Kamioka} = 0.51 \pm 0.07.
\end{equation}
The errors in (1) -- (4) are the added in quadrature statistical and
systematic errors as separately estimated by each collaboration.

 It has been argued\cite{bp},\cite{bbethe},\cite{hatalan0} that the
theoretical uncertainties of the standard solar model alone cannot
account for the discrepancy between theoretical predictions and the
experimental results. However, these uncertainties have to be taken
into account in a conservative analysis of the data, as in some cases
they are bigger than the experimental errors. We include in our
analysis the theoretical uncertainties, as well as the correlations
between the uncertainties in the predicted event rates in the
different solar neutrino detectors, as described in\cite{hatalan} for
an analogous MSW analysis. The uncertainties of the different solar
neutrino fluxes and detection cross--sections estimated in
\cite{bp} have been used. We include an uncertainty of two percents for
the low energy electron--neutrino scattering cross-section which has
not been measured with high enough precision at the neutrino energies
of interest. Finally, we prefer to treat the SAGE and GALLEX results
as independent measurements and do not use the corresponding weighted
average result.

 The probability for an $\nu_e$ to remain $\nu_e$ has been averaged
over a period of one year taking into account the ellipticity of the
Earth orbit, as described in\cite{kp}. Both the change of the survival
probability and the change of the total flux with the distance between
the Sun and the Earth have been included in the calculation.

  The comparison between expected and measured event rates in each
detector has been made for a sufficiently large number of pairs \dm2
and \sn2. For oscillations into active neutrinos,
$\nu_e\leftrightarrow\nu_a$, $\nu_a$ being either $\nu_{\mu}$ or
$\nu_{\tau}$, the minimal $\chi^2$ ($\chi^2_{min}$) is 4.6 with
theoretical uncertainties included in the analysis, and 5.2 if the
theoretical uncertainties are neglected. With four experimental
results (two degrees of freedom) this means that as a solution of the
solar neutrino problem the $\nu_e\leftrightarrow\nu_a$ oscillations
are ruled out at 90\% C.L., but are allowed at 95\% C.L. Accepting the
hypothesis that the $\nu_e\leftrightarrow\nu_a$ oscillations provide
the solution of the solar neutrino problem, the 90\% C.L. and 95\%
C.L. allowed regions of values of \dm2 and \sn2 are shown in
Figs. 1. The results depicted in Fig. 1a have been obtained with the
theoretical uncertainties taken into account as discussed earlier,
while the results shown in Fig. 1b have been derived without including
the theoretical uncertainties in the analysis.

  For solar neutrino oscillations into a sterile neutrino,
$\nu_e\leftrightarrow\nu_s$, the same analysis shows that
$\chi^2_{min} = 10.1$ and $\chi^2_{min} = 11.1$, respecively. Thus,
the $\nu_e\leftrightarrow\nu_s$ oscillations are ruled out at 99\%
C.L. as a solution of the solar neutrino problem \footnote{This result
depends slightly on the threshold used in the Kamiokande detector,
which here was assumed to be 7.5 MeV.}. In case one accepts to combine
the results of the two Ga--Ge detectors and to have only three data
points, the oscillations into sterile neutrinos are excluded at 99.5\%
C.L.
\footnote{Let us note that the MSW $\nu_e$ transitions into a sterile
neutrino $\nu_s$ also give a rather poor fit of the mean event rate
data: in this case $\chi^2_{min} = 3.43$ which implies that the
indicated transitions are ruled out as a solution of the solar
neutrino problem at 80 \% C.L. \cite{kp2}.}

   Let us note that in none of the previous studies
[10] - [13] the constraints on \dm2 and \sn2 following from the mean
event rate data only were derived.

  A second approach to the analysis of the solar neutrino data has
been proposed and described in detail in\cite{kp}. It is more suited
for an analysis of data that is supposed to vary periodically with
time. In such case the time intervals over which the detected signal
is averaged should be chosen shorter than the period of the
anticipated variations in order not to smear out the latter.
Therefore, for neutrino oscillations in vacuum data averaged over
shorter than one year periods should be used and the comparison should
be made on a ``run by run'' basis. This method allows to rule out
certain regions of parameters that are allowed by the analysis which
makes use only of the mean values of the measured event rates.
Although the individual runs have large error bars, the effect of a
systematic discrepancy between predicted and observed event rates
results in a higher $\chi^2$ for certain values of \dm2 and \sn2. On
the other hand, the opposite effect might also occur. When introducing
more degrees of freedom the overall $\chi^2$ might increase slower
than the value of the percentage point, $\chi_p$, for the
corresponding number of degrees of freedom. Therefore values of the
parameters ruled out by the analysis using mean values only, might
become allowed if one utilizes the ``run by run'' data.

  For the analysis of the data from the solar neutrino experiments in
terms of neutrino oscillations in vacuum this approach was first
pursued in\cite{kp} using only the results of the Homestake and
Kamiokande--II experiments. Here the same procedure is applied adding
the additional information about the measured $^{71}{\rm Ge}$
production rate in each of the 30 individual runs completed by the
GALLEX collaboration. In the case of vacuum neutrino oscillations
the variations of the event rates in any solar neutrino detector
resulting from the change of distance between the Sun and the Earth
(apart from the standard geometrical one) are mostly due to $^7{\rm
Be}$ neutrinos. The latter have a very narrow spectrum\cite{bahc3}
which can be approximated by a line. Therefore the averaging over the
continuous spectrum of the other components of the solar neutrino flux
leads to much less pronounced variations of the signals due to these
neutrinos as compared with the signals due to $^7{\rm Be}$ neutrinos
\cite{kp},\cite{kp2} (see also\cite{bacfrau}). In the Homestake
detector the $^7{\rm Be}$ neutrinos contribute only about 14~\% of the
total expected signal, whereas in the Ga--Ge detectors they contribute
$\sim$25~\% of the signal. Therefore, if vacuum neutrino oscillations
are the solution of the solar neutrino problem, the seasonal
variations of the signal in Ga--Ge detectors should be somewhat
stronger than in Cl--Ar ones. As the GALLEX data seems to be rather
constant with time, one expects that certain regions of \dm2 and \sn2
parameters can be ruled out.
However, the results obtained within the more
detailed approach cannot be directly compared with those presented in
Figs. 1 as data for short time intervals
from the Kamiokande experiment for the period since 1990 have not been
published yet. In order to compare the effectiveness of the two approaches
we give below also the constraints following from the mean values of
the measured to expected event rates for the Homestake experiment,
eq. (1), the GALLEX, eq. (3), and the Kamiokande--II experiment
\begin{equation}
R_{\rm Kamioka} = 0.46 \pm 0.08.
\end{equation}
Note also, that we have not included the theoretical uncertainties in
the "run by run" analysis because of the formidable computational
difficulties of inverting large correlation matrices.

   Our results for the case of $\nu_e\leftrightarrow\nu_a$
oscillations are shown in Figs. 2. When only the mean values are taken
into account one obtains $\chi^2_{min} = 2.8$.  For 3 data points (1
degree of freedom) this implies that the hypothesis of
$\nu_e\leftrightarrow\nu_a$ oscillations being the solution of the
solar neutrino problem is ruled out at 90\% C.L. but cannot be ruled
out at 95 \% C.L. Once this hypothesis has been accepted, the regions
of \sn2 and \dm2 allowed at 90\% C.L. and at 95\% C.L. in this case are
shown in Fig. 2a.

 With the run by run event rates used in the analysis we have
$\chi^2_{min} = 117$ for 125 degrees of freedom, which implies a good
quality of the fit. The allowed regions of parameters \dm2 and \sn2 at
90\% C.L. and 95\% C.L. are shown in Fig. 2b. They are considerably
narrower than those depicted in Fig. 2a. The region around \dm2 = $1.1
\times 10^{-10}$ eV$^2$, which is allowed at 90\% C.L. if one uses the
mean values in the analysis, is practically completely ruled out even
at 95 \% C.L. by the run-by-run data.

  The same analysis has been performed for the hypothesis of solar
neutrino oscillations into a sterile neutrino,
$\nu_e\leftrightarrow\nu_s$. With only the mean values used in the
analysis the minimal \ch2 is equal to 6.7. Consequently, the
$\nu_e\leftrightarrow\nu_s$ oscillations give a poor fit of the mean
event rate data (1), (3) and (5): they are marginally allowed only at
99.5\% C.L. The situation is quite different when the variations are
taken into account. In this case $\chi^2_{min} = 123$, which means
that the $\nu_e\leftrightarrow\nu_s$ oscillations are now allowed even at
68\% C.L. The corresponding 90\% C.L. and 95\% C.L. allowed
regions of \dm2 and \sn2 are shown in Fig. 3. It should be noted that
the latest GALLEX-I results not only do not constrain further the
allowed values of the parameters \dm2 and
\sn2, but actually slightly relaxes the constraints obtained from the
Homestake and Kamiokande--II data only. Thus, we see that the mean
event rate data are much more restrictive in the case of vacuum
oscillations into sterile neutrino (to the point of practically
excluding them as a possible solution of the solar neutrino problem),
than the run-by-run data. The inverse is true for the
$\nu_e\leftrightarrow\nu_{\mu,\tau}$ oscillations: the stronger
restrictions on the parameters follow from the run-by-run results.

  In conclusion, the solution of the solar neutrino problem in terms
of two-neutrino oscillations in vacuum has been confronted with the
latest data from all solar neutrino experiments. It has been shown
that previously allowed regions of the parameters \sn2 and \dm2 are ruled
out by the data from individual runs of the Homestake and GALLEX
experiments, and data from the Kamiokande-II detector, averaged over
three--month periods. The $\chi^2-$analysis of the current mean event
rates in the Homestake, Kamiokande, SAGE and GALLEX detectors based on
the Bahcall -- Pinsonneault theoretical predictions rules out
two-neutrino $\nu_e$ oscillations into sterile states as a solution of
the solar neutrino problem at 99\% C.L., with the uncertainties in the
theoretical predictions included in the analysis.

\vskip 0.5cm
   We would like to thank J. Bahcall for very useful correspondence.
The kind hospitality and partial support of the Institute for Nuclear Theory
at the University of Washington, where the present Updated Version of our
earlier work has been completed, is acknowledged with gratefulness.
\vskip 0.5cm

{\bf Note Added.} The present Updated Version of our preprint IFP - 480 - UNC,
Ref. SISSA 177/93/EP, differs from the one published in November 1993 by the
data from the GALLEX collaboration used as input in the analyses: here we
utilize the latest GALLEX results published in February of 1994, which are
based on data from 30 completed runs of measurements (see ref. [12]); in the
version of our work from November 1993 only the data from the first 15
acomplished runs of the GALLEX experiment, available at that time, were used.
As a consequence, the results presented here differ somewhat from the results
published in November 1993.

\newpage
\centerline{\bf Figure Captions}
\medskip
\noindent
{\bf Fig. 1} Regions of values of the parameters \dm2 and \sn2 allowed
at 90 \% C.L. (solid line) and 95 \% C.L. (dashed line) in the case of
vacuum $\nu_e\leftrightarrow\nu_{\mu(\tau)}$ oscillations of solar
neutrinos.  The mean event rates measured by the Homestake,
Kamoikande, SAGE and GALLEX collaborations, and the SSM predictions of
Bahcall and Pinsonneault have been used in the $\chi^2-$analysis. The
results shown have been obtained a) by including, and b) without
including, the theoretical uncertainties in the analysis.

\medskip

\noindent
{\bf Fig. 2} Regions of values of the solar neutrino
$\nu_e\leftrightarrow\nu_{\mu(\tau)}$ oscillation parameters
\dm2 and \sn2 allowed at 90 \% C.L. (solid line) and 95 \% C.L. (dashed line).
Fig. 2a has been obtained by using only mean event rates (eqs. (1),
(3), and (5), see the text). The results of the extended
\ch2--analysis based on 84 runs of the Homestake experiment, 30 runs
of the GALLEX experiment and 13 three--month time--intervals of the
Kamiokande-II experiment are given in Fig. 2b.

\medskip

\noindent
{\bf Fig. 3} The same as in Fig. 2b for solar neutrino oscillations
into sterile neutrino in vacuum: $\nu_e\leftrightarrow\nu_s$. The
black dots (seen at \dm2 $\cong 7.9\times 10^{-11}~eV^2$ and \sn2 $\cong
0.79$) correspond to values of \dm2 and \sn2 allowed at 95\% C.L.;
they also give an idea about the precision of the calculations.
\newpage
\input epsf
\hsize=10cm
\epsfbox{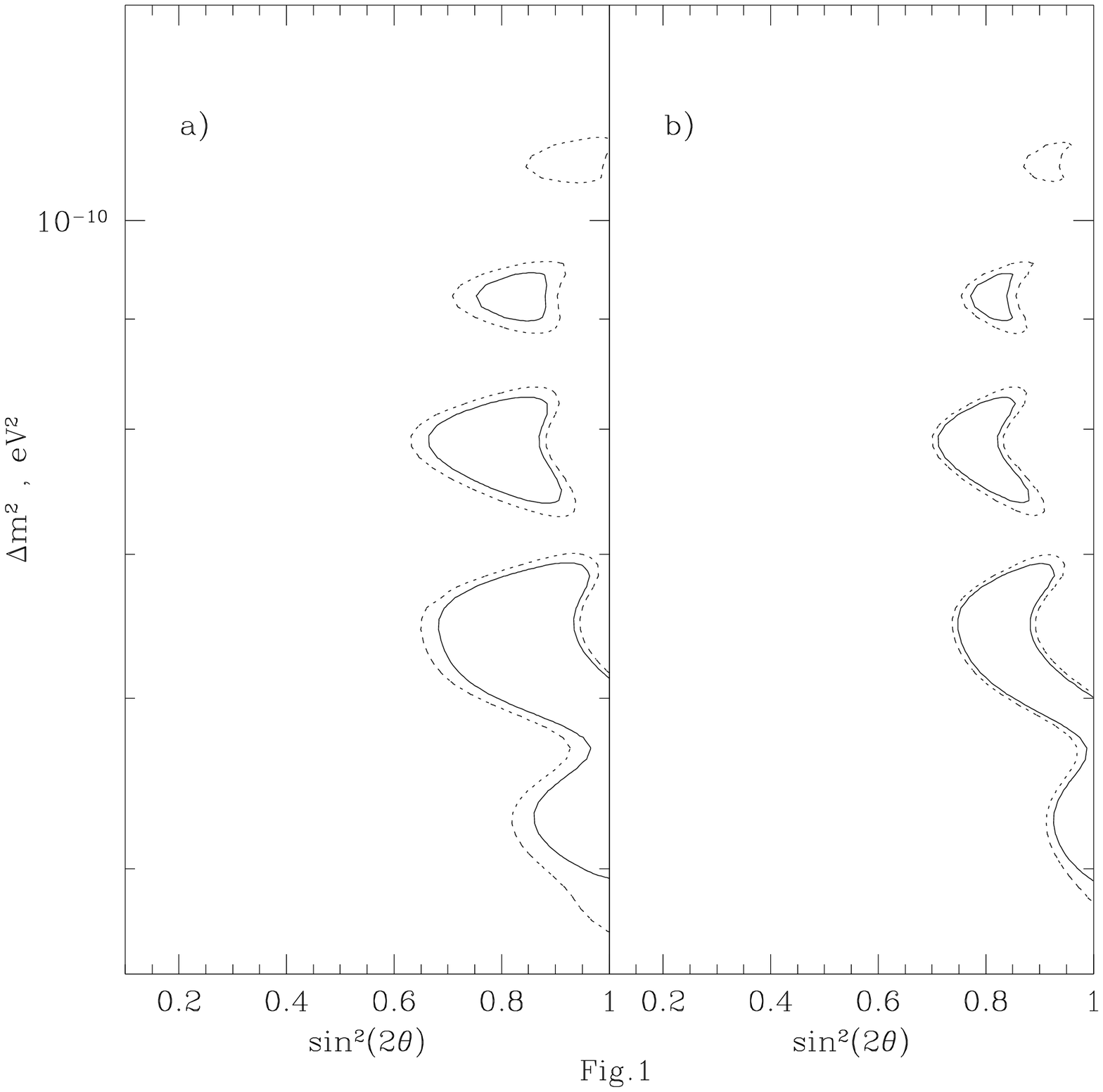}
\newpage
\epsfbox{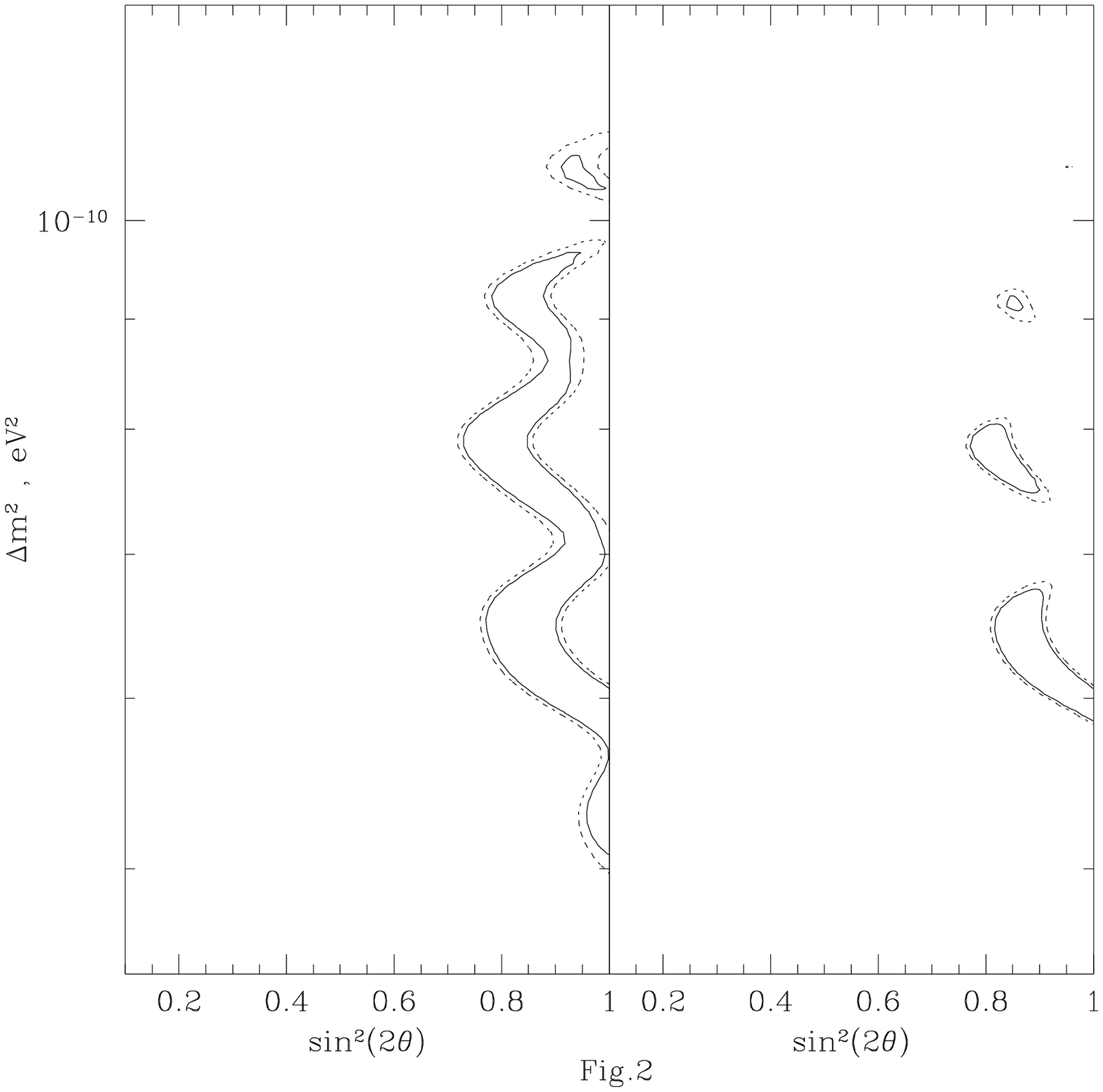}
\newpage
\epsfbox{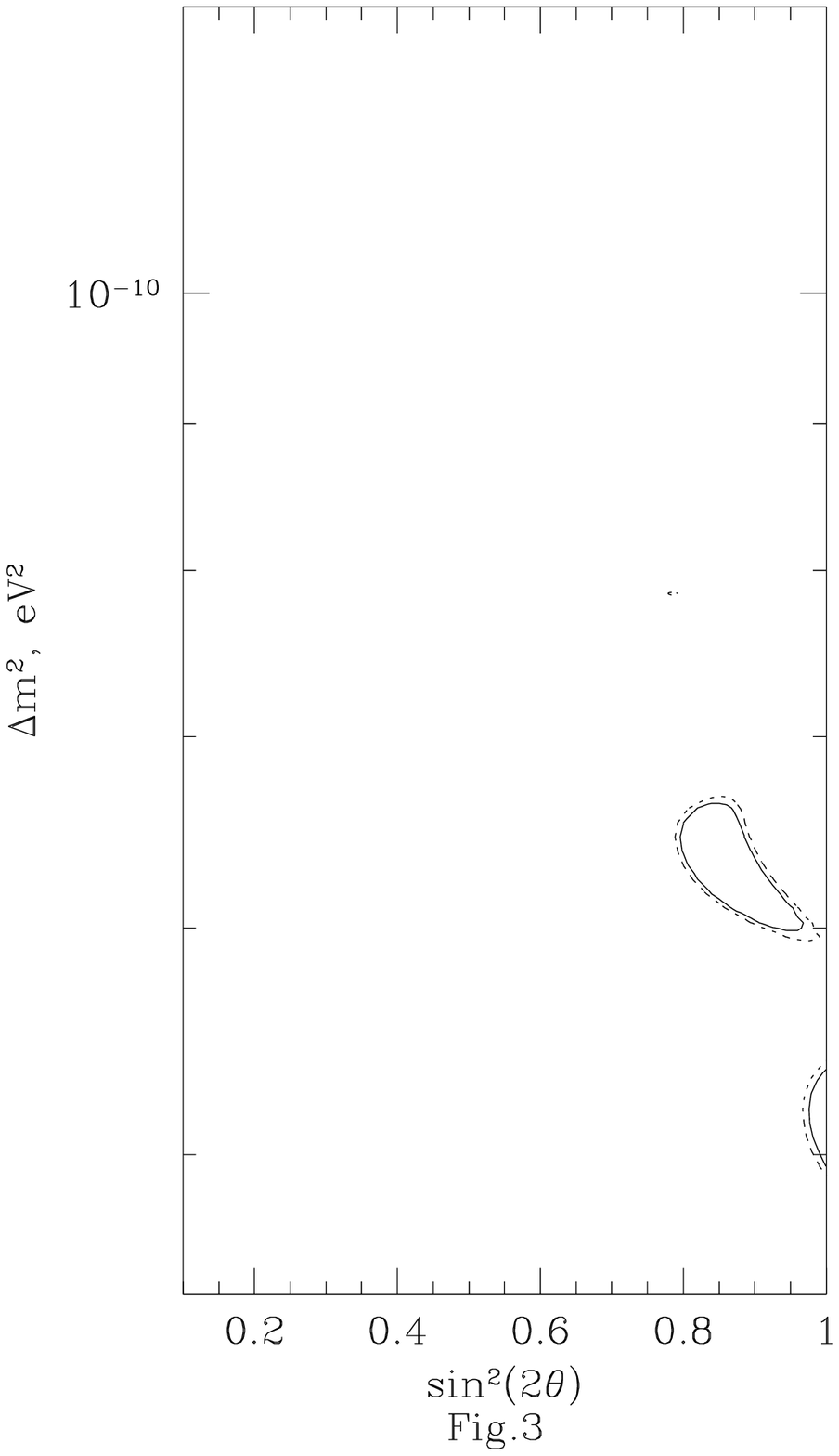}
\end{document}